\documentclass[preprint2]{emulateapj}
\usepackage{txfonts}
\usepackage{graphicx,subfigure}
\usepackage{epsfig}

\usepackage{rotating}

\newcommand{\mic}{\,{\rm \mu m} } 

\begin{document}
\title{Spitzer characterization of dust in the ionized medium of the Large Magellanic Cloud}
\author{D\'eborah Paradis\altaffilmark{1}, Roberta
  Paladini\altaffilmark{1}, Alberto Noriega-Crespo\altaffilmark{1},
  Guilaine Lagache\altaffilmark{2}, Akiko Kawamura \altaffilmark{3},
Toshikazu Onishi \altaffilmark{4} and Yasuo Fukui \altaffilmark{3}}
\altaffiltext{1}{Spitzer Science Center, California Institute of Technology,Pasadena, CA 91125}
\altaffiltext{2}{Institut d'Astrophysique Spatiale, 91405 Orsay, France}
\altaffiltext{3}{Department of Astrophysics, Nagoya University,
  Chikusa-ku, Nagoya 464-8602, Japan}
\altaffiltext{4}{Department of Physical Science, Osaka Prefecture
  University, Gakuen 1-1, Sakai, Osaka 599-8531, Japan}
\begin{abstract}
A systematic investigation of dust emission associated with the ionized gas has so far been performed only in
our Galaxy and for wavelengths longer than 60 $\mic$. Newly available Spitzer data now offer the opportunity to carry out a
similar analysis in the Large Magellanic Cloud (LMC). 

By cross-correlating Spitzer SAGE (Surveying the Agents of a
Galaxy's Evolution) data with the ATCA/Parkes HI 21-cm 
data, the NANTEN $\rm ^{12}CO$ (J=1-0) data, and both the SHASSA H$\alpha$ and the 
Parkes 6-cm data, we investigate the physical properties of dust associated with 
the different phases of the gas (atomic, molecular and ionized).  
In particular, we study the presence and nature of dust from 3.6 to 160 $\mic$ and for various regimes of the ionized gas, 
spanning emission measures (EM) from $\sim$ 1 pc cm$^{-6}$ (diffuse
component) to $\sim$ 10$^{3}$ pc cm$^{-6}$ (HII regions).

Using a dust emission model, and testing our results with several radiation
field spectra, we show that dust in the ionized gas is warmer
than dust associated with other phases (atomic and
molecular). We also find a decrease of the polycyclic
aromatic hydrocarbons (PAH) relative abundance with respect to big
grains (BGs), as well as an increase of the near infrared (NIR)
continuum. These three results (e.g. warmer temperature, decrease of PAH abundance and increase of the NIR continuum) are found consistently for all regimes of the ionized
gas. On the contrary, the molecular phase appears to provide favorable conditions for the survival of PAHs. Furthermore, 
the very small grain (VSG) relative abundance tends to increase in the ionized phase,
especially in bright HII regions. Last but not least, our analysis shows that the emissivity of dust associated with the ionized gas is lower in the LMC 
than in our Galaxy, and that this difference is not accounted for by the lower metallicity of the LMC.

\end{abstract}  

\section{Introduction}
\label{sec_intro}

Detecting dust emission associated with the diffuse ionized phase is a complex task due to
two main competing factors. First of all, large grains are expected to be selectively
destroyed in the warm ionized medium (WIM) due to the passage of multiple
shock fronts which would reduce their emission. In addition,  along most line
of sights (LOS), the
diffuse ionized gas is likely mixed with the other phases of the gas (atomic and molecular), 
and its contribution is not expected to be dominant at any latitude.

Because of these difficulties, contradictory results concerning the presence of dust in the WIM have been reported in the literature.
\citet{Lagache99}, looking at the high Galactic latitude residuals of the IR/HI correlation between 100 and 1000 $\mic$,
claim a statistical detection. In particular, by adopting an emissivity spectral index of
2, they find an emissivity law in the atomic and ionized gas phase of, respectively, $\rm \tau/N_H=8.7\pm0.9\,10^{-26} \,(\lambda
/250\,\mic)^{-2}$ cm$^{-2}$  and $\rm
\tau/N_H=1.0\pm0.2\,10^{-25} \,(\lambda /250\,\mic)^{-2}$ cm$^{-2}$, with T$_{HI}$ = 17.5 K and T$_{H+}$ = 20 K. 
Following this preliminary investigation, \citet{Lagache00} extend their analysis, and carry out a decomposition of the
far infrared (FIR) emission   
using WHAM \citep{Reynolds98} H$\alpha$ data as a tracer of the ionized medium. The result of this study 
confirms the early detection, with an emissivity
law for the ionized phase of $\rm \tau/N_H=1.1\,10^{-25} \,(\lambda /250\,\mic)^{-2}$
cm$^{-2}$, for a fixed temperature of 17.2 K for both the atomic and ionized phase. At low Galactic latitudes, 
\citet{Paladini07} derive dust emissivities associated with the atomic, molecular and ionized gas phases, 
using an inversion method applied to IRAS (Neugebauer et al. 1984), DIRBE (Hauser 1993), Archeops (Benoit et al. 2002) and WMAP 
(Bennett et al. 2003) data. According to their analysis, the average dust temperatures in the atomic and molecular phases are comparable, and of order of
19.8 and 19.2 K respectively, while dust in the ionized phase is significantly warmer, T$_{H+}$ $\simeq$ 26.7 K.  
At the same time, \citet{Odegard07} looking at the DIRBE/WHAM correlation in five selected 
high-latitude Galactic regions find, contrary to \citet{Lagache00},
that the H$\alpha$-correlated dust component is negligible in all cases.

Despite the observational caveats, assessing the presence of dust in the ionized gas phase is
very important, given that the optical properties of dust grains can vary with the properties of the local
environment. To date, no clear-cut indication of such a behaviour has been found though. 
\citet{Paradis10} perform a statistical study of all the molecular clouds in the
LMC and do not find any systematic change in dust abundances nor in dust temperature between 
the molecular and the atomic phase. On the contrary, \citet{Stepnik03} and \citet{Paradis09b}, 
by looking at nearby Galactic clouds, do report evidence of this kind of variations.  
In the ionized medium, due to the proximity of the ionized gas with UV sources such as O and B stars and the cumulative effect of shock fronts, 
changes in the optical dust properties are indeed expected. For instance, depletion of 
PAHs has
been observed toward individual HII regions \citep{Everett10, Watson09, Contursi07, Peeters05, Povich07}. 
However, this phenomenon does not appear to be systematic, and its origin is still
poorly understood.

In this paper, we present a systematic investigation of the properties of dust associated with the ionized gas in the LMC. 
The LMC, located at a distance of $\sim$ 50 kpc \citep{Feast99,Leeuwen07}, has an advantageous almost face-on viewing
angle. At -32$\degr$ latitude, the contamination along the LOS is expected to be small compared to the plane of the Galaxy.
For this galaxy, \citet{Bernard08} carried out a cross-correlation study of the combined Spitzer SAGE \citep{Meixner06}
and IRAS data, with HI and CO data. That work allowed to
characterize (in terms of emissivities, abundances, etc.) the properties of dust mixed with the atomic and molecular medium, as well as
to point out the existence of an excess of emission at 70 $\mic$. Interestingly, \citet{Bernard08} showed that this excess appears to be
spatially correlated with both the atomic and the diffuse (I$_{H_{\alpha}} <$ 200 R) ionized gas.

The present study builds over and expands the analysis described in \citet{Bernard08}, by 
including the ionized medium in the cross-correlation and by  
encompassing different physical regimes of this phase of the gas, from the most tenuous environments (diffuse medium), to the
densest ones (HII regions).

The paper is organized as follows. Section \ref{sec_data} summarizes
the data we used. In Section \ref{sec_decomp}, we describe the
decomposition method as well as the dust emission modeling. We present our main results in Section \ref{sec_res}. Concluding remarks
are provided in Section \ref{sec_cl}.

\section{Observations}
\label{sec_data}
\subsection{IR data}
Along the lines of \citet{Bernard08}, we use, as a tracer of dust emission from the NIR to
the FIR, the Spitzer SAGE data. The data span the wavelength range 3.6
$\mic$ to 160 $\mic$, with an angular resolution from $\simeq\,2
^{\prime \prime}$ to 40$^{\prime \prime}$, respectively. 

These Spitzer data are combined with the 12 and 100 $\mic$ IRIS data \citep[Improved Reprocessing
of the IRAS Survey,][]{MamD05} characterized by an angular resolution close to
4$^{\prime}$. The inclusion of the IRAC 8$\mic$ and IRIS 12 $\mic$ data allows us to take into account the contribution from the most prominent PAH features, i.e. at 7.7 and 11.3 $\mic$, while with MIPS 24 and 70 $\mic$ the bulk of the emission from VSGs is incorporated. 
In addition, the 100 $\mic$ IRIS data point, combined with the MIPS 160 $\mic$ measurement, 
provides the necessary leverage for the derivation of the equilibrium
temperature of the BGs.

\subsection{Gas tracers}
\subsubsection{HI data}
As a tracer of atomic gas, we use the \citet{Kim03} 21-cm map,  which is a combination of
interferometric data obtained with the Australia Telescope Compact
Array (ATCA), at 1$^{\prime}$ angular resolution, and the Parkes antenna, at 15.3$^{\prime}$ angular
resolution \citep{Staveleysmith03}. 
Assuming the gas is optically thin, the column density can be derived 
using the relation:
\begin{equation}
\label{eq_xhi}
N_{HI}=X_{HI}W_{HI}
\end{equation}
where $\rm W_{HI}$ is the HI integrated intensity map and the conversion
factor $\rm X_{HI}$ is taken equal to $1.82\times10^{18}$ $\rm H/cm^2/(K\,km/s)$ \citep{Spitzer78}. The integrated map 
was built by summing up all the velocity channels from 190 to 386 km s$^{-1}$. For the purpose of estimating and subtracting 
the Galactic foreground contribution to each IR wavelength, we use a Galactic 
HI column density map constructed by \citet{Staveleysmith03}. More details on this procedure can be found in \citet{Bernard08}.

\subsubsection{CO data}
The $^{12}$CO (J=1-0) survey obtained with the 4-m radio NANTEN
telescope of Nagoya University at Las Campanas Observatory, Chile
\citep[see][]{Fukui08} is adopted as a tracer of the molecular
gas. In particular, we make use of the second NANTEN survey which, characterized by an angular resolution of
2.6$^{\prime}$\footnote{about 40 pc at the distance of the LMC}, 
evidenced the existence in that galaxy of 272 individual molecular clouds \citep{Fukui08}. 

We compute the molecular column densities using the relation: 
\begin{equation}
N_{H_2}=X_{CO}W_{CO}
\end{equation}
where $\rm W_{CO}$ is the CO integrated intensity map, generated by integrating the original CO cube over the 
200 $<$ v$\rm _{lsr} <$ 320 km s$^{-1}$ velocity range. \citet{Fukui08}
derived the $\rm X_{CO}$ conversion factor assuming that each cloud is at virial equilibrium. They found an average value
for all the clouds of $\rm X_{CO}=(7 \pm 2) \times 10^{20}$
H$_2$/cm$^2$/(K km s$^{-1}$). From cloud to cloud, values vary
from 1.6$\times10^{20}$ to 4.2$\times10^{21}$ $\rm
H_2/cm^2/(K\,km/s)$. Their average value is larger than that 
estimated by \citet{Hughes10} of $4.7\times 10^{20}$  $\rm
H_2/cm^2/(K\,km/s)$, using the Mopra telescope data, at a resolution of 33$^{\prime
  \prime}$. This difference is due to the fact that low CO brightness
clouds have been excluded from the MOPRA survey, and the $\rm
X_{CO}$ value has been deduced from a smaller number of clouds
compared to NANTEN observations. In this work, in order to be
consistent with our previous studies \citep[e.g.][]{Paradis10}, 
we use the $\rm X_{CO}$ values determined for
each individual cloud by \citet{Fukui08}. For clouds for which the virial mass could not be
derived, we adopt the LMC average value of $\rm X_{CO}$=$(7 \pm 2) \times
10^{20}$ H$_2$/cm$^2$/(K km s$^{-1}$). For a comprehensive tabulation of $\rm X_{CO}$ values, we 
refer the reader to \citet{Paradis10} (see their Table 5).

\subsubsection{H${\alpha}$ and radio data}
\label{sec_halpha_tracers}
The warm ionized gas, characterized by an electron temperature of the order of $\rm T_e \simeq 10^4$ K, emits both 
recombination lines and radio free-free
continuum. Among the observed recombination lines, the H$\alpha$ line is the brightest and 
is due to a transition of the Balmer line series at 6562.81 $\AA$. As for free-free emission, this   
consists of free electrons being scattered and decelerated by their encounters with ions.  
H$\alpha$ and radio templates are available for the LMC and both tracers are used in our analysis to assure an 
independent cross-check of the results. 
For the H$\alpha$, we use the
Southern H-Alpha Sky Survey Atlas \citep[SHASSA,][]{Gaustad01} data set which
covers the southern hemisphere ($\delta <15\degr$), at an angular
resolution of $0.8^{\prime}$, and with a sensitivity of 2 Rayleigh/pixel (1
R=10$^6$ photons cm$^{-2}$ s$^{-1}$). Following \citet{Lagache99}, if
we assume that the electron density $\rm n_e$ is constant along each LOS,
the $\rm H^+$ column density can be derived using the relation: 
\begin{equation}
\label{eq_halpha}
\frac{N_H^{H^+}}{H\,cm^{-2}}=1.37 \times 10^{18} \frac{I_{H
    \alpha}}{R} \frac{n_e}{cm^{-3}}
\end{equation}
where 1 R=2.25 pc cm$^{-6}$ for T$\rm _e$=8000 K
\citep[see, for instance][]{Dickinson03}. For $\rm n_e$,
we adopt different values for different regimes of the ionized
gas, as described in Section \ref{sec_reg}. 

To trace the radio free-free emission, we use observations obtained with the Parkes 64-m
telescope at 4.75 GHz \citep[see][for a complete
description of the Parkes survey]{Filipovic95}. At that frequency the
beam size is 4.9$^{\prime}$, but after correcting for scanning
effects, the effective beam size results to be 
equal to $5.6^{\prime}$. 

In order to compare the radio and H$\alpha$ maps and compute gas column densities for the radio map, we
transform the radio emission to H$\alpha$ units, using Equation 11 in \citet{Dickinson03}:
\begin{equation}
\label{eq_tb}
\frac{T_b}{I_{H \alpha}}=8.396 \times 10^3 a \times \nu_{GHz}^{-2.1}T_4^{0.667}10^{0.029/T_4}(1+0.08)
\end{equation}
where $\rm T_b$ (brightness temperature) is in $\rm \mu K$ and $\rm I_{H
  \alpha}$ in Rayleigh. $\rm T_4$ is 
$\rm T_e$ in units of $10^4$ K, while the factor $a$ depends on the
frequency and the electron temperature \citep[see Table 3
in][]{Dickinson03} and $\rm \nu_{GHz}$ is the frequency in GHz. For
$\rm T_e=$ 8000 K ($\rm T_4=0.8$), $a$ is close to unity. To apply Equation
\ref{eq_tb}, we first convert the original units of the Parkes map
(mJy/beam) into Jy, and these into T$\rm _b$ using the expression:
\begin{equation}
T_b=\frac{S_{4.75GHz}c^2}{2k\nu^2}
\end{equation}
with $\rm S_{4.75GHz}$ the total flux density at 4.75 GHz, c the speed of light, $\nu$ the frequency and k the Boltzmann
constant. Once the Parkes map has been converted into $\rm I_{H \alpha}$ units, 
the H$^+$ column density can be computed by applying Equation \ref{eq_halpha}. 
The sensitivity of the Parkes telescope is 8 mJy/beam, also equivalent to 28 R, i.e. 
significantly lower than the sensitivity threshold of the SHASSA map. This implies, as we will also see in Section 3.2, that while 
the SHASSA map will be used as a tracer of the ionized gas for all types of environmental conditions (i.e. from 
the diffuse to the dense ionized medium), the Parkes map will only be
used to probe the densest/brightest regions. 

For the pixels corresponding to bright HII regions, we have performed a comparison 
of the radio Parkes map, converted into H$\alpha$ emission ($\rm I_{H\alpha}$), with the SHASSA map. The 
comparison highlights that, for 88$\%$ of the pixels, the H$\alpha$-Parkes data 
are higher than the SHASSA one. The median value of the ratio between the two data sets 
is 1.6. To explain such a discrepancy, we invoke extinction as a source of attenuation in the 
SHASSA map. In this respect, we remind the reader that H$\alpha$ measurements are affected by dust absorption, 
while radio data provide an uncontaminated observational window on the ionized gas.  
In the LMC, extinction is usually considered rather low. For 
instance, \citet{Imara07} and \citet{Dobashi08} quote an average A(V) of 0.3 mag. 
In our analysis, the factor 1.6 of discrepancy would induce an
extinction value in the SHASSA map of 0.51 mag in bright HII
regions. Following \citet{ODonnell94}, \citet{Dickinson03} estimate
the absorption in H$\alpha$ as $A(H\alpha)=0.81A(V)$. This result is
obtained using an extinction curve for the  Milky Way. However, as
shown by \citet{Gordon03}, the extinction curve in the LMC is very
similar to the Galactic one. Therefore, the estimate above also
applies here. In that case, the corresponding extinction of LMC bright
HII regions in the visible is A(V)=0.63 mag. This value is slightly higher than
the extinction estimated using the LMC A(V) map
constructed by \citet{Dobashi08}, and weighted by the H$\alpha$
emission ($\sum A(V) I_{H\alpha}/\sum I_{H\alpha} \simeq 0.57$ mag) over the same HII
regions.  It is unlikely that bright HII
regions are all confined behind dust clouds, which may indicate that the
extinction could have been underestimated when constructing the LMC A(V)
map from star counts, as already suggested in \citet{Paradis10}. In
the following, we therefore expect differences of about a factor 1.6 when comparing
results using SHASSA and Parkes data (see Section \ref{sec_case3_results}).

\vspace*{0.7truecm} 
All the data used for the analysis (IR, HI, CO, H$\alpha$) have been degraded to 
the lowest angular resolution of the Parkes 4.75-GHz map, 
using a Gaussian kernel with a fwhm of 5.6$^{\prime}$. After being convolved, 
the maps have been reprojected onto the same grid, with a 2$^{\prime}$ pixel size. Finally, point source
subtraction has been applied to all the IR maps \citep[see][for the
description of the map processing]{Bernard08}.

\begin{sidewaystable*}[!h]
\caption{Results of the IR emission decomposition with respect to different
gas phases, in unit of MJy/sr for 10$^{20}$ H/cm$^2$, and their
1$\sigma$ uncertainties. The different
regimes of the ionized gas described in Section \ref{sec_reg} are denoted  -- in the first column -- with 1 (diffuse ionized gas), 2
(typical HII regions) or 3 (bright HII
regions; 3a-using the H$\alpha$ data and 3b-using the radio data at
4.75 GHz).\label{table_val}}
\begin{center}
\resizebox{\textwidth}{!}{%
\begin{tabular}{lccccccccc}
\hline
\hline
Cases & I$_{160} $ & I$_{100}$  & I$_{70}$ & I$_{24}$ & I$_{12}$
&I$_{8}$  &  I$_{5.8}$  & I$_{4.5}$  & I$_{3.6}$   \\
\hline
1-HI & (5.57$\pm$0.03)$\times 10^{-1}$ & (3.19$\pm$0.03)$\times
10^{-1}$ & (1.61$\pm$0.02)$\times 10^{-1}$ & (5.48$\pm$0.07)$\times
10^{-3}$ & (7.65$\pm$0.08)$\times 10^{-3}$ & (10.27$\pm$0.09)$\times
10^{-3}$ & (2.05$\pm$0.46)$\times 10^{-4}$ & (6.14$\pm$0.24)$\times
10^{-4}$ & (1.03$\pm$0.04)$\times 10^{-3}$ \\
1-CO & (4.02$\pm$0.02)$\times 10^{-1}$ & (1.79$\pm$0.02)$\times
10^{-1}$ & (6.77$\pm$1.42)$\times 10^{-2}$ & (5.38$\pm$0.45)$\times
10^{-3}$ & (1.04$\pm$0.05)$\times 10^{-2}$ & (1.08$\pm$0.06)$\times
10^{-2}$ & (5.48$\pm$0.29)$\times 10^{-3}$ & (3.19$\pm$1.55)$\times
10^{-4}$ & (8.58$\pm$2.49)$\times 10^{-4}$ \\
1-H$\alpha$ & (5.36$\pm$0.02)$\times 10^{-1}$ & (4.01$\pm$0.22)$\times
10^{-1}$ & (3.72$\pm$0.18)$\times 10^{-1}$ & (1.06$\pm$0.06)$\times
10^{-2}$ & (4.44$\pm$0.67)$\times 10^{-3}$ & (9.54$\pm$0.69)$\times
10^{-3}$ & (2.26$\pm$0.42)$\times 10^{-3}$ & (3.35$\pm$0.21)$\times
10^{-3}$ & (5.27$\pm$0.32)$\times 10^{-3}$ \\
\hline
2-HI & (6.03$\pm$0.03)$\times 10^{-1}$ & (3.34$\pm$0.03)$\times
10^{-1}$ & (2.38$\pm$0.02)$\times 10^{-1}$ & (7.31$\pm$0.07)$\times
10^{-3}$  & (7.37$\pm$0.07)$\times 10^{-3}$   & (9.92$\pm$0.08)$\times 10^{-3}$ &  (1.07$\pm$0.04)$\times
10^{-3}$ & (3.36$\pm$0.17)$\times 10^{-4}$ & (5.67$\pm$0.27)$\times 10^{-4}$ \\
 2-CO &  (7.52$\pm$0.11)$\times 10^{-1}$ & (4.79$\pm$0.11)$\times
 10^{-1}$ & (2.17$\pm$0.08)$\times 10^{-1}$ & (1.47$\pm$0.03)$\times
 10^{-2}$ & (2.12$\pm$0.03)$\times 10^{-2}$ &  (2.13$\pm$0.04)$\times
 10^{-2}$ & (7.89$\pm$0.18)$\times 10^{-3}$ & (1.51$\pm$0.08)$\times
 10^{-3}$ & (2.46$\pm$0.12)$\times 10^{-3}$   \\
2-H$\alpha$ & 3.69$\pm$0.06 & 3.88$\pm$0.06 & 2.83$\pm$0.05 &
(1.32$\pm$0.02)$\times 10^{-1}$ &  (6.32$\pm$0.18)$\times 10^{-2}$ &
(7.16$\pm$0.20)$\times 10^{-2}$ &  (2.20$\pm$0.10)$\times 10^{-2}$ &
(9.81$\pm$0.42)$\times 10^{-3}$ & (1.37$\pm$0.07)$\times 10^{-2}$ \\
\hline
3a-HI & (8.80$\pm$0.17)$\times 10^{-1}$ & (5.68$\pm$0.02)$\times
10^{-1}$ & (4.62$\pm$0.10)$\times 10^{-1}$ & (2.21$\pm$0.07)$\times
10^{-2}$ & (7.49$\pm$0.44)$\times 10^{-3}$ & (1.41$\pm$0.03)$\times
10^{-2}$ & (2.30$\pm$0.16)$\times 10^{-3}$ & (4.50$\pm$0.36)$\times
10^{-4}$ & (7.00$\pm$0.57)$\times 10^{-4}$ \\
3a-CO & 1.65$\pm$ 0.05 & 1.78$\pm$0.07 & (5.50$\pm$0.30)$\times
10^{-1}$ & (3.28$\pm$0.22)$\times 10^{-2}$ & (4.55$\pm$0.15)$\times
10^{-2}$ & (3.59$\pm$0.11)$\times 10^{-2}$ & (1.62$\pm$0.05)$\times
10^{-2}$ & (1.92$\pm$0.12)$\times 10^{-3}$ & (3.01$\pm$0.18)$\times
10^{-3}$ \\
3a-H$\alpha$ & 8.30$\pm$0.22 & 11.19$\pm$0.29 & 5.43$\pm$0.13 &
(6.18$\pm$0.09)$\times 10^{-1}$ & (1.80$\pm$0.06)$\times 10^{-1}$ &
(1.16$\pm$0.04)$\times 10^{-1}$ & (4.84$\pm$0.21)$\times 10^{-2}$ &
(1.77$\pm$0.05)$\times 10^{-2}$ & (1.81$\pm$0.07)$\times 10^{-2}$ \\
\hline
3b-HI & (7.32$\pm$0.18)$\times 10^{-1}$ & (3.71$\pm$0.23)$\times
10^{-1}$ & (3.67$\pm$0.11)$\times 10^{-1}$ & (1.34$\pm$0.09)$\times
10^{-2}$ & (5.14$\pm$0.51)$\times 10^{-3}$ & (1.23$\pm$0.04)$\times
10^{-2}$ & (1.88$\pm$0.19)$\times 10^{-3}$ & (2.23$\pm$0.41)$\times
10^{-4}$ & (4.03$\pm$0.61)$\times 10^{-4}$ \\
3b-CO & 1.47$\pm$0.05 & 1.55$\pm$0.07 & (4.28$\pm$0.30)$\times
10^{-1}$ & (2.19$\pm$0.25)$\times 10^{-2}$ & (4.29$\pm$0.15)$\times
10^{-2}$ & (3.39$\pm$0.11)$\times 10^{-2}$ & (1.56$\pm$0.06)$\times
10^{-2}$ & (1.63$\pm$0.12)$\times 10^{-3}$ & (2.66$\pm$0.18)$\times
10^{-3}$ \\
3b-radio & 4.98$\pm$0.12 & 6.74$\pm$0.16 & 3.22$\pm$0.07 &
(3.40$\pm$0.06)$\times 10^{-1}$ & (9.92$\pm$0.36)$\times 10^{-2}$ &
(6.67$\pm$0.26)$\times 10^{-2}$ & (2.39$\pm$0.13)$\times 10^{-2}$ &
(9.73$\pm$0.28)$\times 10^{-3}$ & (1.08$\pm$0.04)$\times 10^{-2}$ \\
\hline
\end{tabular}}
\end{center}
\end{sidewaystable*}

\section{Decomposition of the IR emission}
\label{sec_decomp}
\subsection{Regimes of the ionized gas \label{sec_reg}}

We consider three regimes of the ionized gas, i.e. from diffuse H$^{+}$ to bright HII regions, and 
for each regime we investigate the dust properties independently, 
due to potentially intrinsic variations in dust composition, evolution and physics of the grains. To define each regime, 
we set three thresholds in H$\alpha$ emission intensity using the SHASSA  map, convolved to 5.6$^{\prime}$ resolution, and the catalog of HII regions in the LMC compiled by \citet{Kennicutt86} (hereafter KH86). 

\begin{figure*}[h]
\begin{center}
\includegraphics[width=8.5cm]{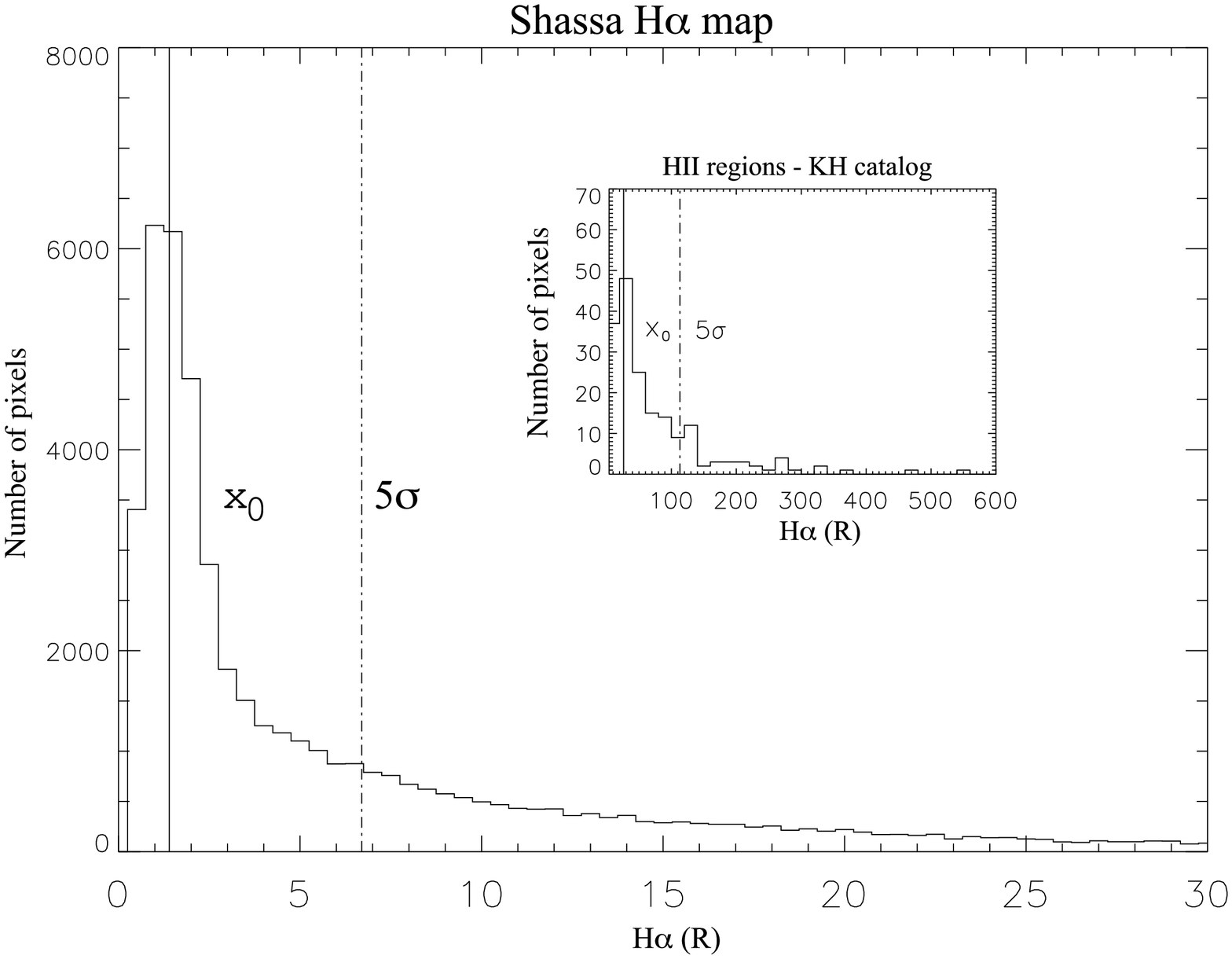}
\caption{Pixel intensity distribution of the SHASSA H$\alpha$ map. Values on the x-axis have been 
truncated to 30 R to evidence the bulk of the distribution. Overlaid are: (solid line) the mean (x$_{0}$) and 
(dashed line) 5$\sigma$ values. Insert panel: pixel intensity distribution for the KH86 HII regions in the SHASSA map. 
The mean value and 5$\sigma$ limits are also shown.  \label{fig_ne}}
\end{center}
\end{figure*}

\begin{figure*}
\begin{center}
\includegraphics[width=16cm]{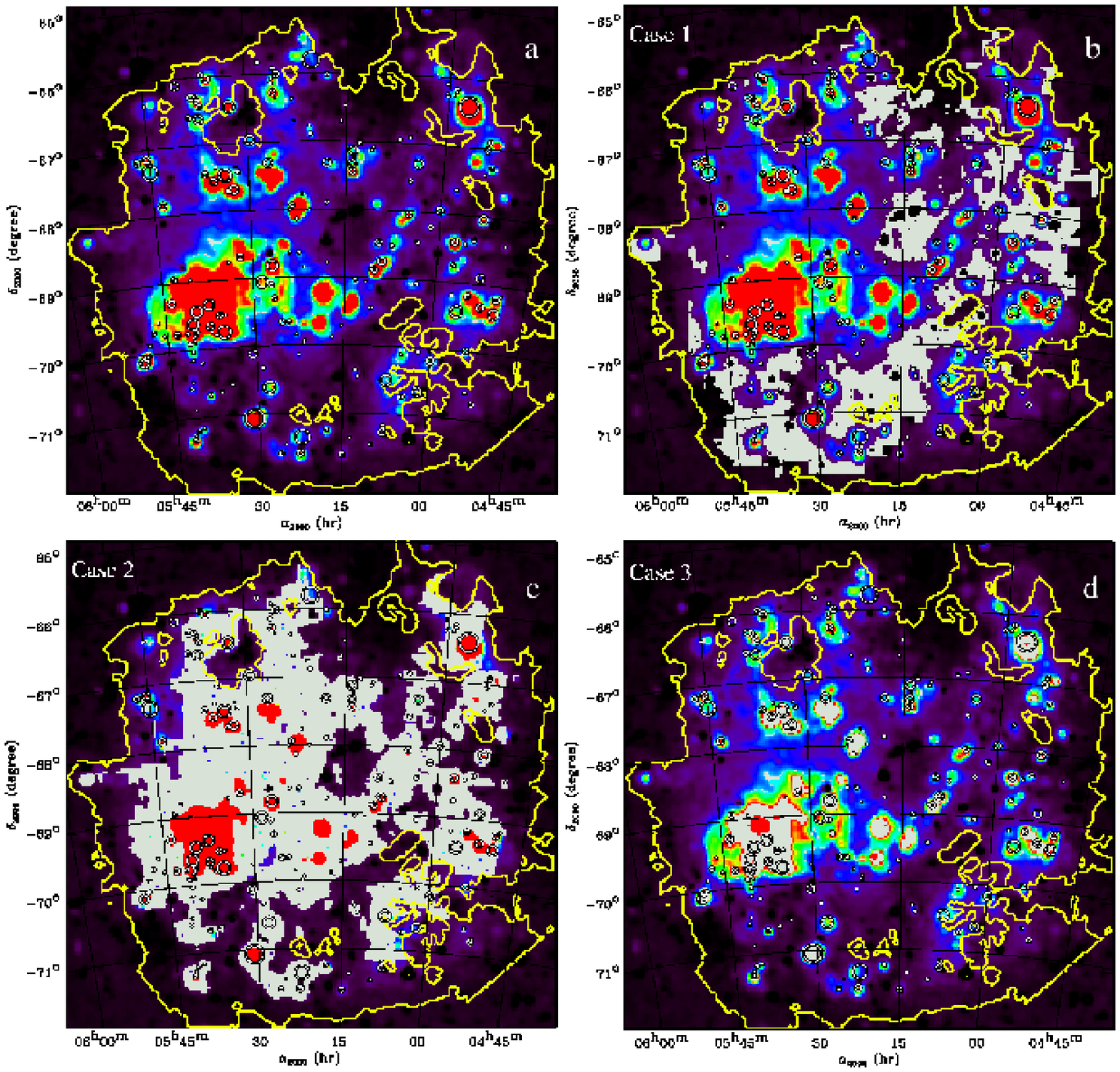}
\caption{SHASSA H$\alpha$ map (a) and spatial locations of pixels (in grey) selected for 
  each case: (b)- case 1 (diffuse ionized gas), (c)- case 2
  (typical HII regions), and (d)- case 3 (bright HII regions).
  The overlaid symbols show the HII regions \citep{Kennicutt86}. The
  range of the maps is [-8 ; 100] R.\label{fig_cases}}
\end{center}
\end{figure*}

We start by analyzing the pixel intensity distribution in the SHASSA map (see Figure \ref{fig_ne}). The mean 
($\rm \overline{I_{H\alpha}}$) of the distribution is 1.4 R, with a
dispersion ($\sigma$) of 1.1 R and a minimum of 0.4 R, which is the
sensitivity of the SHASSA data for the 5.6$^{\prime}$ angular
resolution and the 2$^{\prime}$ pixel size of the map. 
About 99$\%$ of the pixels lies in the range 0.4 R $\rm < I_{H\alpha} <$ 6.7 R,  
corresponding to $\rm \overline{I_{H\alpha}} \pm$ 5$\sigma$. 
We notice that none of the sources in the KH86 catalog are located in regions 
of the SHASSA map where the H$\alpha$ intensity is within this range (see Figure \ref{fig_cases}), and therefore, we adopt it to define the {\em{diffuse}} ionized gas regime. 

For $\rm I_{H\alpha}>$ 6.7 R, we look, in the SHASSA map, at the
intensity distribution of the pixels which correspond to the
  location of the HII regions in the KH86 list. To select the pixels, we use, for each cataloged source, its coordinates and measured angular 
diameter. The H$\alpha$ intensity distribution obtained by this process is shown in Figure \ref{fig_ne} (insert panel). This is 
characterized by a mean value of 26.3 R, with a dispersion of 17.4 R.  
In this case, 99$\%$ of the pixels lies in the range 6.7 R $\rm < I_{H\alpha} <$ 113.3 R, and such a range is 
used to define what, in this context, we denote as {\em{typical}} HII regions. 

For $\rm I_{H\alpha} >$  113.3 R, we have a few very bright HII regions which are listed in the KH86 catalog. We set an upper limit of 1000 R 
to exclude extremely bright sources, such as 30 Doradus, which -- due to their complexity - would require a dedicated analysis. 
The range 113.3 R $\rm < I_{H\alpha} <$ 1000 R hence identifies the third regime.   

In summary, we have three distinct regimes/cases which we will be tackled by our analysis:

\begin{itemize}
\item{case 1:  0.4 R $\rm < I_{H\alpha} <$ 6.7 R (diffuse ionized gas);}
\item{case 2: 6.7 R $\rm < I_{H\alpha} <$ 113.3 R (typical HII regions);}
\item{case 3: 113.3 R $\rm < I_{H\alpha} <$ 1000 R (very bright HII regions).}
\end{itemize}

For each regime, we  need to adopt a characteristic $\rm n_{e}$ in
order to compute the H$\rm ^{+}$ 
column density from Equation \ref{eq_halpha}. For the diffuse component, no estimate 
is available from the literature. We have made an attempt to derive this value directly from 
the SHASSA map. For this purpose, we first compute, for each pixel in the
range 0.4 R $\rm < I_{H\alpha} <$ 6.7 R,  
the emission measure (EM), by applying Equation \ref{eq_tb} in
\citet{Dickinson03} and assuming T$\rm _{e}$ = 8000 K. 
To derive the rms electron density from the average EM so computed
($\rm \overline{EM}$ = 5.3 $\pm$ 0.1 pc cm$^{-6}$), we need to assume a linear scale 
of the emitting ionized medium. Our 5.6$^{\prime}$-SHASSA map  has a pixel size of 2$^{\prime}$. At 
a distance of 50 kpc, this corresponds to a linear size of $\sim$ 30 pc. However, such a value 
is very likely a severe underestimate of the actual vertical extent of the diffuse ionized gas. In fact, KH86 have 
analyzed the distribution in linear diameters for their cataloged HII regions, and obtained 
that the largest sources in the sample have a size of about 400 pc. In the hypothesis that the diffuse 
gas extends far beyond the layer of discrete HII regions, if we denote
with $\rm d_{H^{+}}$ the vertical extent 
of H$\rm ^{+}$ and take $\rm d_{H^{+}}$ = 500 pc, we obtain an average electron
density n$\rm _{e_{1}}$ of 0.10 $\pm$ 0.08 cm$^{-3}$. This value is
close to what has been derived for our Galaxy, i.e. n$\rm _{e}$ $\sim$ 0.03 -- 0.08 cm$^{-3}$, for 
an estimated scale height ranging between 1000 and 1800 pc \citep{Haffner09}. If, instead of 500 pc, we assume 
$\rm d_{H^{+}}$ = 1000 pc, we retrieve n$\rm _{e_{1}}$ = 0.07 $\pm$ 0.06 cm$^{-3}$, matching the Galactic value. 
Despite such a result, given the uncertainty in $\rm d_{H^{+}}$ for the LMC, we choose  
to adopt the average value derived from Galactic
measurements. Therefore, for case 1, we take n$\rm _{e_{1}}$  = 0.055 cm$^{-3}$. 
 
For cases 2 and 3, we compute EM, for each HII region in the KH86 catalog by applying Equation 4 in 
\citet{Dickinson03}. For each source, we use the quoted H$\alpha$ flux and 
we set T$\rm _{e}$ = 8000 K. The measured  H$\alpha$ flux is converted into Rayleighs, and the intensity cuts 
at 113.3 R and 1000 R for, respectively, case 2 and 3 are applied. The averaged emission measures for the two 
cases ($\rm \overline{EM_{2}}$ = 183.8 $\pm$ 17.8  pc cm$^{-6}$,
$\rm \overline{EM_{3}}$ = 1322.4 $\pm$ 167.7  pc cm$^{-6}$) 
are then used to derive an rms electron density. Following 
this procedure, we obtain: n$\rm _{e_{2}}$ = 1.52 $\pm$ 0.44 cm$^{-3}$ and
n$\rm _{e_{3}}$ = 3.98 $\pm$ 2.05 cm$^{-3}$, i.e. 20 and 80 times larger
than the diffuse component.
 
\begin{table*}
\caption{Results of the modeling obtained by assuming a single radiation field along the LOS. Three different regimes of the
  ionized gas are considered: diffuse (1), typical HII regions (2) and bright
  HII regions (3a corresponds to H$\alpha$ data and 3b to radio data
 at 4.75 GHz). The first column denotes the adopted radiation field spectrum:
  the \emph{Mathis} RF or the \emph{4-Myr cluster} Galev RF (see Section \ref{sec_mod_single}).\label{table_cases}}
\begin{center}
\begin{tabular}{lcccccc}
\hline
\hline
RF &Cases & $\rm
Y_{tot}$ (D/G) & NIR cont & $\rm X_{RF}$ &  $\rm Y_{PAH}/Y_{BG}$ & $\rm
Y_{VSG}/Y_{BG}$\\
& &  $10^{-3}$ & $10^{-4}$ & &$10^{-2}$  & $10^{-2}$ \\ 
\hline
Mathis &1-HI & 1.82$\pm$0.04 & 1.98$\pm$0.22 & 1.51$\pm$0.03 & 4.29$\pm$0.20 & 5.37$\pm$0.28 \\
Mathis  &1-CO &  3.19$\pm$0.87 & 4.90$\pm$1.36 & 0.61$\pm$0.16 & 9.87$\pm$5.19 & 6.65$\pm$3.78 \\
Mathis &1-H$\alpha$ &  0.50$\pm$0.07& 36.71$\pm$1.83 & 6.46$\pm$0.90 &1.50$\pm$0.54 & 9.47$\pm$2.80 \\
\hline 
Mathis  & 2-HI & 1.62$\pm$0.03 & 1.02$\pm$0.15 & 1.86$\pm$0.03 &4.00$\pm$0.14 & 7.08$\pm$0.26 \\
Mathis   & 2-CO &  2.77$\pm$0.17 & 14.44$\pm$0.69 & 1.42$\pm$0.08 & 8.94$\pm$1.07 & 10.55$\pm$1.35 \\
4-Myr cluster & 2-H$\alpha$ & 3.16$\pm$0.15 & 91.69$\pm$3.83 &0.39$\pm$0.02 & 0.88$\pm$0.09 & 6.88$\pm$0.66 \\
\hline
Mathis & 3a-HI &  1.29$\pm$0.08 & 3.06$\pm$0.33 & 3.93$\pm$0.24 & 3.27$\pm$0.41 & 13.81$\pm$1.90 \\
Mathis & 3a-CO &  3.61$\pm$0.35 & 16.47$\pm$1.07 & 2.79$\pm$0.26 & 6.38$\pm$1.20 & 9.13$\pm$2.10\\
4-Myr cluster &3a-H$\alpha$ & 12.67$\pm$0.94 & 157.06$\pm$4.41 & 0.21$\pm$ 0.02 & 0.77$\pm$0.12 & 17.35$\pm$2.61 \\
\hline
Mathis & 3b-HI &  1.10$\pm$0.10 & 0.78$\pm$0.37 & 3.66$\pm$0.30 & 3.54$\pm$0.59 & 10.28$\pm$2.08 \\
Mathis & 3b-CO &  3.34$\pm$0.36 & 13.77$\pm$1.12 & 2.65$\pm$0.28 & 6.83$\pm$1.45 & 6.26$\pm$1.95 \\
4-Myr cluster & 3b-Radio (5GHz) & 7.33$\pm$0.51 & 87.89$\pm$2.65 & 0.22$\pm$0.01 & 0.68$\pm$0.10 & 15.48$\pm$2.20 \\
\hline
\end{tabular}
\end{center}
\end{table*}

\begin{table*}[!t]
\caption{Results of the modeling obtained by assuming a composite radiation field along the LOS. Three different regimes of the
  ionized gas are considered: diffuse (1), typical HII regions (2) and bright
  HII regions (3a corresponds to H$\alpha$ data and 3b to radio data
  at 4.75 GHz). The first column denotes the adopted composite
  radiation field spectrum:
  the \emph{Mathis} RF or the \emph{4-Myr cluster} Galev RF (see Section \ref{sec_mod_mixture})\label{table_cases_sum}}
\begin{center}
\begin{tabular}{lccccc}
\hline
\hline
RF &Cases &  $\rm
Y_{tot}$ (D/G) & $\alpha$ &  $\rm Y_{PAH}/Y_{BG}$ & $\rm Y_{VSG}/Y_{BG}$\\
& & $10^{-3}$ &  &$10^{-2}$  & $10^{-2}$ \\ 
\hline
$\rm \sum_i \sum_j Mathis $ &1-HI & 3.39$\pm$0.02  & 2.47 &4.57$\pm$0.06 & 2.11$\pm$0.08 \\
$\rm \sum_i \sum_j Mathis$ & 1-CO &  2.35$\pm$0.13 & 2.50 & 8.68$\pm$0.78 & 4.17$\pm$0.91 \\
$\rm \sum_i \sum_j Mathis$ & 1-H$\alpha$ & 3.06$\pm$0.12& 2.29 & 2.74$\pm$0.28 &0.43$\pm$0.50 \\
\hline 
$\rm \sum_i \sum_j Mathis+4$-$\rm Myr\,cluster$ & 2-HI & 6.23$\pm$0.03 & 2.50 & 2.42$\pm$0.03 &1.56$\pm$0.05 \\
$\rm \sum_i \sum_j Mathis+4$-$\rm Myr\,cluster$ & 2-CO &  7.83$\pm$0.11 & 2.50 & 4.91$\pm$0.12 & 3.93$\pm$0.22 \\
$\rm \sum_i \sum_j 4$-$\rm Myr\,cluster$ & 2-H$\alpha$ & 1.81$\pm$0.02 & 2.50 &0.88$\pm$0.03 & 1.02$\pm$0.09 \\
\hline
$\rm \sum_i \sum_j Mathis+4$-$\rm Myr\,cluster$ & 3a-HI &  8.38$\pm$0.16 & 2.32 & 1.66$\pm$0.07 & 0.27$\pm$0.26 \\
$\rm \sum_i \sum_j Mathis+4$-$\rm Myr\,cluster$ & 3a-CO & 19.38$\pm$0.58 & 2.50 & 3.38$\pm$0.17 & 3.81$\pm$0.55 \\
$\rm \sum_i \sum_j 4$-$\rm Myr\,cluster$ & 3a-H$\alpha$ & 3.53$\pm$0.08 & 2.50 & 0.76$\pm$0.04 & 8.39$\pm$0.39 \\
\hline
$\rm \sum_i \sum_j Mathis+4$-$\rm Myr\,cluster$ & 3b-HI &  7.17$\pm$0.19 & 2.38 & 1.89$\pm$0.10 & 0.72$\pm$0.41 \\
$\rm \sum_i \sum_j Mathis+4$-$\rm Myr\,cluster$ & 3b-CO & 17.01$\pm$0.59 & 2.50 & 3.61$\pm$0.20 & 2.08$\pm$0.64 \\
$\rm \sum_i \sum_j 4$-$\rm Myr\,cluster$ & 3b-Radio (5GHz) & 2.16$\pm$0.04 & 2.50 & 0.71$\pm$0.04 & 7.00$\pm$0.36 \\
\hline
\end{tabular}
\end{center}
\end{table*}

\subsection{Method}
\label{sec_method}
In the LMC, \citet{Bernard08} evidenced the existence of a large FIR excess
 with respect to the atomic, molecular and ionized gas phases, which they attributed
to either molecular gas with no associated CO emission or to optically thick HI. They
showed that this component is spatially correlated with
HI. \citet{Paradis10} performed a statistical analysis of LMC molecular
clouds and showed that if the FIR excess was not taken into account in
the IR correlations, the dust abundance in the atomic phase would be
systematically overestimated by a factor 1.7 (see their Section 3). Following
\citet{Paradis10}, we decompose the IR emission at each wavelength and for each regime, as a function of the atomic, molecular and ionized gas column density:
\begin{equation}
\label{eq_gen}
I(\lambda)=a(N_H^{HI}+N_H^X)+bN_H^{CO}+cN_H^{H+}+d
\end{equation}
with $\rm N_H^X$ the column density in the FIR excess component. In
the following, we refer to the sum of the atomic and FIR excess
components as the atomic phase. Despite this working hypothesis, it is important to keep in mind that the FIR excess could be
molecular in nature, instead of atomic, and associated with warm H$_2$ gas \citep{Planck11}.
The parameters a, b, c, 
denoting the emissivities associated with each phase of the gas at a given wavelength, and the constant d are determined using regression fits. 
To remove any possible offsets in the data, we subtract
a background in all the maps.
The background is
computed as the median over a common area, corresponding to
the faintest 10$\%$ of the HI data. This step also ensures to have a null IR emission for a null column density. The H$\alpha$ data are
used for all regimes of ionized gas described in the previous
section, while, due to the limited sensitivity of the 4.75-GHz map, we
can only use the radio data for case 3. As a consequence, a direct cross-check of the results 
obtained by adopting both tracers will only be possible for this case.

Although case 3 corresponds to bright HII regions for which, presumably, most of the radio emission is due to thermal free-free, yet we might have a residual non-thermal contamination associated with synchrotron radiation.
To investigate the synchrotron contamination,
we complement the 4.75-GHz Parkes data with the 1.4-GHz data, obtained with the same telescope. Both maps are smoothed to the common
resolution of 16.6$^{\prime}$, characterizing the 1.4-GHz data. Following \citet{Niklas97}, we estimate the fraction of emission contributed by free-free as: 
\begin{equation}
\label{eq_sync}
\frac{S_{4.75}}{S_{1.4}}=f_{th} \left ( \frac{\nu_{4.75}}{\nu_{1.4}}
\right ) ^{-\alpha_{ff}}+ (1-f_{th}) \left ( \frac{\nu_{4.75}}{\nu_{1.4}}
\right ) ^{-\alpha_{sync}}
\end{equation}
where $\rm f_{th}$ is the thermal fraction at 4.75
GHz, $\rm S_{4.75}$ and $\rm S_{1.4}$ are the total flux densities at 4.75 and 1.4
GHz, and $\rm \alpha_{ff}$ and $\rm \alpha_{sync}$ are the
free-free and synchrotron spectral indices. Following \citet{Paladini05} and \citet{Hughes06}, we take 
$\rm \alpha_{ff}$ = 0.1. The spectral index of synchrotron radiation is known to present large spatial variations. 
In our Galaxy, it ranges between 0.6 in star-forming complexes to 1.2 in the interarm regions. If, in Equation \ref{eq_sync}, we adopt a fixed value of $\rm \alpha_{sync}$=0.6, we derive an average thermal fraction in the LMC of 0.85. This value is probably an underestimate due to the fact
that, while $\rm \alpha_{ff}=0.1$ is a good approximation for
frequencies less than or equal to a few GHz and $\rm T_e=8000\,K$, the actual free-free spectral index is  
a weak function of both frequency and electron temperature \citep{Bennett92}. Moreover, HII regions span a wide range of T$\rm _e$ values. If we compare our result with similar attempts found in the literature, we see that, for instance, \citet{Meinert93} tried to
separate the thermal and non thermal component in the radio emission
of the LMC, using H$\alpha$ and FIR data, at a resolution of
51$^{\prime}$. At 4.75 GHz, they obtained a thermal fraction of
0.59. However their result applies to both the diffuse medium and discrete HII regions, while we 
only consider HII regions.\\
In order to take the synchrotron contamination into account, we multiply the 4.75-GHz
radio map by the derived thermal fraction equal to 0.85.\\
Results of the correlations for each phase and regime of the gas are
presented in Figure \ref{fig_allcases} and in Table \ref{table_val}.

\subsection{Modeling}
\label{sec_mod}
\subsubsection{Single radiation field along the LOS}
\label{sec_mod_single}
\begin{figure*}
\begin{center}
\includegraphics[width=8.5cm]{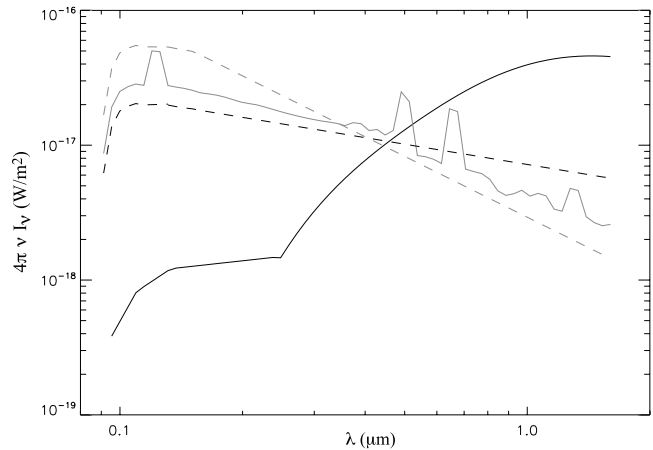}
\caption{Radiation field templates used in this work: \emph{Mathis} (solid black
  line), \emph{4-Myr cluster} Galev (solid grey line), 
and \emph{4-Myr cluster} Galev (gas lines removed) with a hardness
proportional to $\lambda^{-0.5}$ and $\lambda^{-1.5}$ (dashed black
and dashed grey lines). \label{fig_isrf}}
\end{center}
\end{figure*}

From Equation \ref{eq_gen} we obtain the emissivities associated with each phase of the gas for each wavelength $\lambda$ 
in the 3.6 $\mic$ - 160$\mic$ range, and with these values we build the corresponding spectral energy distributions (SEDs). 
The SEDs are modeled with \emph{DustEM} \citep{Compiegne08, Compiegne10}. The \emph{DustEM} code is an updated version of the
\citet{Desert90} model. The code allows to constrain mass abundances ($\rm Y_{BG}$, $\rm Y_{VSG}$, and $\rm
Y_{PAH}$), relative to hydrogen, of the
three dust populations. The intensity of a NIR continuum, with a temperature of
1000 K, is also a free parameter of the model. Indeed, the inclusion of this component is required, as discussed in several studies \citep[e.g.][]{Lu03}, and 
allows us to reproduce observational data at 3.6 and 4.5 $\mic$.  
Two different radiation fields (RFs) input models are used in this work. The strength of the field ($\rm
X_{RF}$) is a free parameter in both cases. However, for all 3 cases (diffuse medium, typical and very bright HII regions), we
impose, for dust emission associated with the
atomic and molecular phases, as well
as for dust emission associated with the diffuse ionized gas in case
1, the same interstellar RF hardness as for the solar
neighborhood \citep{Mathis83}. In the following we will refer to this
RF as \emph{Mathis} RF. We note that, for dust in the atomic and
molecular gas phases, we assume that the dominant component of the RF is always the one related to the interstellar solar neighborhood.
However for cases 2 and 3, dust in the ionized phase is associated with typical and bright
HII regions, and therefore it is essentially heated by
local hot stars. In these regions, the solar neighborhood interstellar RF is
not a good approximation. Following
\citet{Kawamura09}, the youngest stellar objects are less than 10
Myr. Using the GALEV code\footnote{available at http://www.galev.org},
which is an implementation of evolutionary synthesis models, we have generated an UV/visible
spectrum of young stellar clusters of 4 Myr. In the following, we will refer to this RF as the
\emph{4-Myr cluster} RF. The summary of the RFs used for each case and phase
is described in Table \ref{table_cases}, and the RF spectral shapes are shown in Figure \ref{fig_isrf}. 

\subsubsection{Composite radiation fields along the LOS}
\label{sec_mod_mixture}
Assuming a single RF component along the LOS is a simplistic
hypothesis. Therefore we tested our results using instead a combination
of several RF intensities and hardnesses along the LOS. For case 1, we
apply the ``local-SED combination'' model proposed by \citet{Dale01} for normal star forming galaxies. 
This model assumes a
power-law distribution of dust mass subject to a given heating
intensity $\rm dM_d(X_{RF})$:
\begin{equation}
dM_d(X_{RF}) \propto X_{RF}^{-\alpha}dX_{RF},
\end{equation}
where $\rm \alpha$ controls the relative contribution of the various
strengths of the field to the SEDs. \citet{Dale01} considered
RF strengths and $\rm \alpha$ values in the range $\rm
0.3<X_{RF}<10^5$ and $\rm 1<\alpha<2.5$, respectively. A value of
$\alpha$ close to 1 indicates an active star-forming region, whereas $\alpha$ around 2.5 corresponds to the diffuse medium. 
This parameter is linked to the I$\rm _{60}$/I$_{100}$ ratio, and therefore gives an indirect measure of dust temperature.\\ 
Here is how we proceed in our analysis. We first compute emission
spectra using \emph{DustEM}, for different $\rm X_{RF}$ values, and the \emph{Mathis} RF (noted
$\rm I^{mod}_{\nu}(X_{RF},RF_{\odot})$). Then we sum up these emission spectra over the same $\rm
X_{RF}$ range as proposed by \citet{Dale01}, according to:
\begin{equation} \label{eq_isrf}
I^{tot}_{\nu}=\frac{{\sum_i\sum_j}
  I^{mod}_{\nu}(X_{RF,i},RF_{\odot}) \times
  X_{RF,i}^{-\alpha _j}}{{\sum_i\sum_j} X_{RF,i}^{-\alpha _j}}.
\end{equation}

For the atomic and molecular phases in cases 2 and 3, we check the
impact of adopting a combination of the \emph{Mathis} RF and the simulated \emph{4-Myr cluster} Galev RF. Equation \ref{eq_isrf} becomes:

\begin{eqnarray}
\label{eq_isrf_mix}
I^{tot}_{\nu}&=&\frac{\frac{1}{2} \sum_i\sum_j
    I^{mod}_{\nu}(X_{RF,i},RF_{\odot}) \times X_{RF,i}^{-\alpha_j}
  }{\sum_i\sum_j X_{RF,i}^{-\alpha_j}} \\ \nonumber
&+&\frac{\frac{1}{2} f\sum_i\sum_j
    I^{mod}_{\nu}(X_{RF,i},RF_{Galev}) \times X_{RF,i}^{-\alpha_j} }{\sum_i\sum_j X_{RF,i}^{-\alpha_j}}
\end{eqnarray}

The coefficient $f$ allows to normalize the emission spectra computed
with each RF hardness. Indeed, the same $\rm X_{RF}$ will not give
the same absolute emission level depending on the RF used. To
determine $f$, we adjust the emission spectra using the two RFs (\emph{Mathis} RF and \emph{4-Myr cluster} RF) and look for the
coefficient allowing the same brightness at the maximum emission. This
coefficient $f$ is re-derived for each $\rm X_{RF}$. 
For the ionized phase in cases 2 and 3, we apply 
Equation \ref{eq_isrf}, but replacing the \emph{Mathis} RF by the \emph{4-Myr cluster} RF. 

The \emph{DustEM} code does not allow to perform a model fit using a combination of
RFs. Therefore, we have to use a different method to analyze the
effect of a composite RF. Following \citet{Paradis09a}, in order to derive mass
abundances for each dust grain species, we pre-compute the IR
brightnesses associated with each gas phase, using the Galactic
abundances obtained by \citet{Desert90}, for different $\rm \alpha$
values. We then
compare the observed brightnesses between 8 and 160 $\mic$ with the predicted values, and
infer the dust mass abundances minimizing the $\chi^2$ for each value of
$\alpha$. We select the solution with the lowest $\chi^2$. The
additional NIR
continuum is not taken into account to compute the IR brightnesses,
since we do not know the origin of this component, nor how to treat it
in the case of a combination of RFs. For this reason, we do not
consider IR data below 8 $\mic$. The summary of the RF combinations
used for each case and phase is given in Table \ref{table_cases_sum}. 

\begin{figure*}[!t]
\begin{center}
\includegraphics[width=18cm]{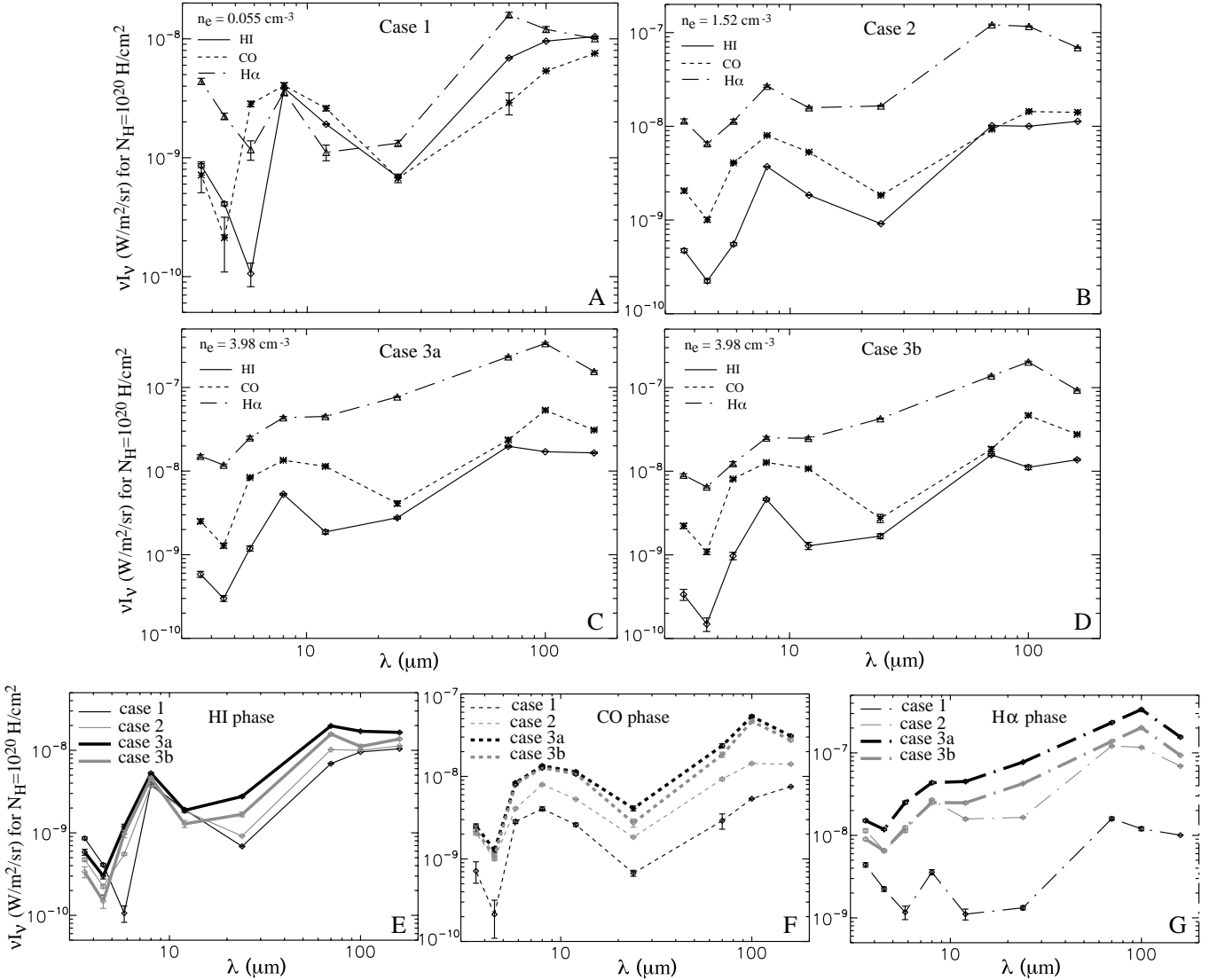}
\caption{Spectral energy distribution associated with the atomic (solid line), 
molecular (dashed line) and ionized (dashed-dotted line)
  phase of the gas, for case 1 (diffuse ionized gas, panel A), case 2
  (typical HII regions, panel B) and case 3 (bright HII regions). For case 3, we show the 
results of the analysis obtained by using both H$\alpha$ (case 3a,
panel C) and 4.75-GHz data (case 3b, panel D). Panels E, F and G:
spectral energy distributions of all cases (case 1
in thin black line, case 2 in thin grey line,
case 3a in thick black line and case 3b in thick grey line), plotted at once for each gas component (same 
notation as in panels A, B, C and D).} \label{fig_allcases}
\end{center}
\end{figure*}

\begin{figure*}
\begin{center}
\includegraphics[width=16cm]{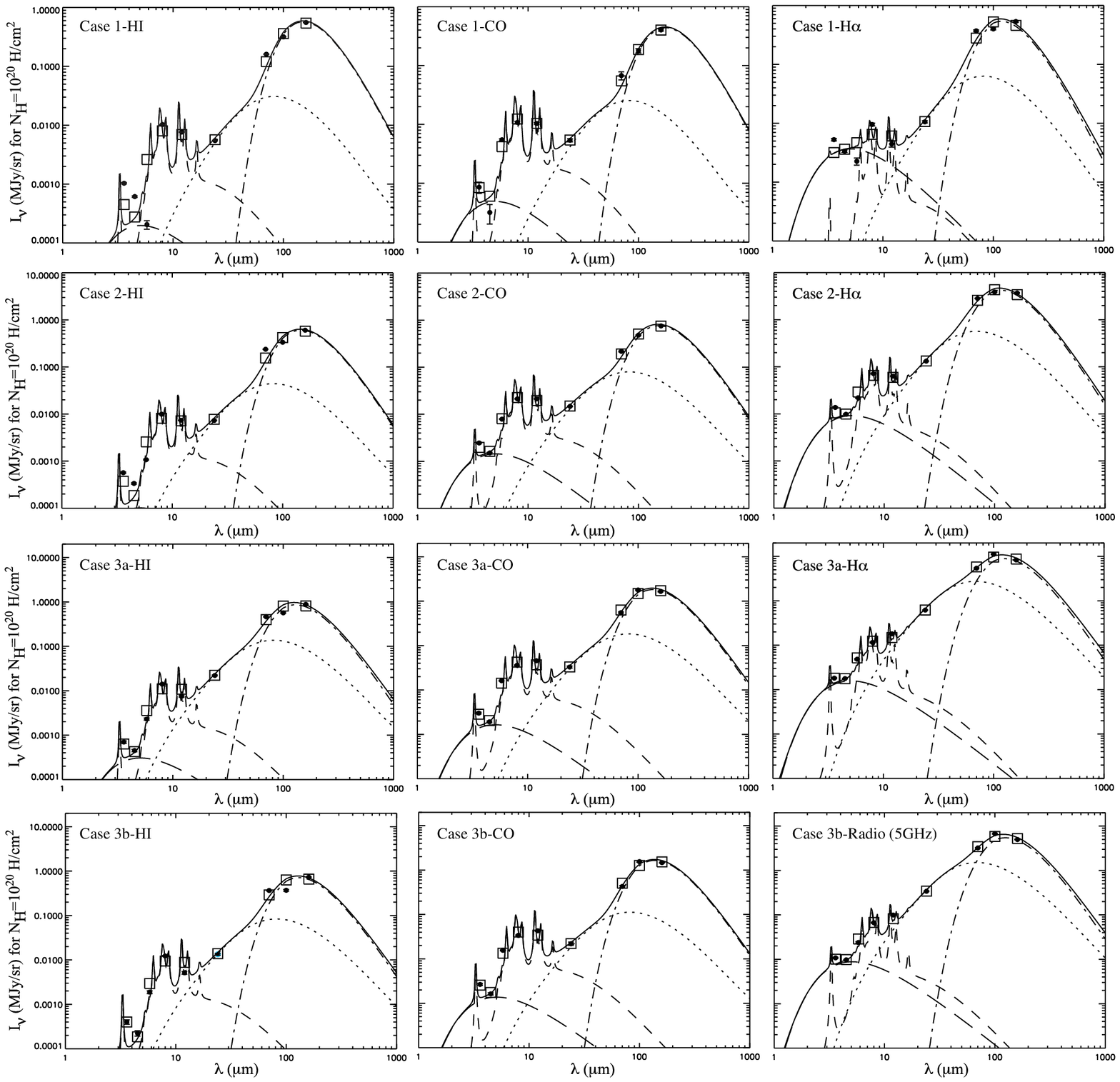}
\caption{Results of the fits obtained with the dust emission model (\emph{DustEM}) assuming a
  single RF along the LOS, for each
  gas phase and for each regime of the  ionized gas : diffuse (case 1),
  typical HII regions (case 2) and bright
  HII regions (case 3). In particular: case 3a corresponds to H$\alpha$ data; case 3b to radio
  data. The squares show the fits of the model after applying the color-correction
  in each Spitzer and IRIS bands. Different curves denote the contributions from various grain populations: total (solid), BG (dot-dash), VSG
  (dot), PAH (dash), and the NIR continuum (long dash).  \label{fig_model}}
\end{center}
\end{figure*}

\section{Results and discussion}
 \label{sec_res}
The dust emission spectra are shown in Figure
\ref{fig_allcases}. Figure \ref{fig_model} illustrates the SED
modeling obtained using \emph{DustEM}, assuming a single radiation
field component. In the following we analyze the
results for the diffuse ionized gas,  typical and 
bright HII regions. We also discuss their consistency 
using the hypothesis of a RFs combination. The best-fit parameters for each 
phase of the gas (molecular, atomic and ionized), and each regime of the 
ionized gas, as well as for different RFs assumptions, are provided in Tables \ref{table_cases} and \ref{table_cases_sum}. 

\subsection{Single radiation field component}
\subsubsection{Case 1: diffuse ionized gas}
\label{subsection_case1}

Figures \ref{fig_allcases} (panel A) and \ref{fig_model} indicate that,
when compared to the atomic and molecular phases, dust emission
associated with the ionized gas phase is a relevant contribution to
the total emission. However, we emphasize that the absolute amplitude of the spectrum for the ionized phase is
    proportional to n$\rm _e$. As a consequence, if n$\rm _e$ has been
    underestimated/overestimated, the absolute level of the spectrum has also been underestimated/overestimated, although the spectral shape will remain unchanged. 
Here, as discussed in Section 3.1, we take n$\rm _e$ equal to  0.055 cm$^{-3}$, which is 
the average observed value in our Galaxy, as reported by \citet{Haffner09}.\\

In this case, the adopted RF hardness is the same for all gas phases, as discussed in Section \ref{sec_mod_single}. 
Therefore, it is possible to perform a direct comparison among the temperature estimates resulting from the best-fit RF intensities. 
We remind the reader that, if we assume a greybody dust emission model with spectral index $\beta$, then the RF intensity and the equilibrium temperature of the BGs are related through:
\begin{equation}
\frac{X_{RF}^{LMC}}{X_{RF}^{\odot}}=\left ( \frac{T_d^{LMC}}{T_d^{\odot}} \right )^{4+\beta}
\end{equation}
where X$\rm _{RF}^{\odot}$ is the RF scaling factor for the \emph{Mathis}
RF (X$\rm _{RF}^{\odot}\simeq 1$), and the dust temperature
in the solar neighborhood $\rm T_d^{\odot}$ is taken equal to 17.5 K
\citep{Boulanger96}, with $\beta=2$. We find dust temperatures equal to 18.7 K, 16.1 K
and 23.9 K for the atomic, molecular and ionized phases,
respectively. These results confirm that dust associated with the
ionized phase is warmer than dust in the atomic and molecular gas
phases, although our derived temperature for the diffuse H$^+$ is slightly lower than 
the value obtained by \citet{Paladini07}, of 26.6 $\pm$ 0.1
K. As for dust emissivity for the ionized phase, we obtain an estimate of 
2.3$\times\,10^{-26}$ cm$^2$/H at 160 $\mic$. For comparison, 
\citet{Lagache00} find, at high latitude, an emissivity of
1.1$\times\,10^{-25}$ cm$^2$/H  at 250 $\mic$,
which corresponds to an emissivity of 2.7$\times\,10^{-25}$ cm$^2$/H
at 160 $\mic$ assuming a spectral index equal to 2, while
\citet{Paladini07}, for the Galactic plane, report a value of  35.1 MJy/sr for 10$^{20}$ H/cm$^2$ 
at 140 $\mic$, corresponding to 8.7 $\times\,10^{-25}$ cm$^2$/H at 160
$\mic$. Therefore, even if in terms of temperature, 
dust in the ionized phase of the LMC is roughly comparable 
to its Galactic plane counterpart, it appears to be 40 times less emissive in the LMC with respect to 
the Galactic plane, and 10 times less emissive with respect to the high Galactic latitudes. 
The low value
in the LMC cannot be only due to the low metallicity of the LMC
\citep[Z $\simeq$ 0.3 - 0.5 Z$_{\odot}$][]{Westerlund97} with respect to our Galaxy, as 
the discrepancy in emissivity is significantly larger than the difference
in metallicity. For the Galactic plane, at least part of the difference could be induced by confusion along the LOS. 
However, we think that different dust properties in the ionized gas phase of the
LMC compared to our Galaxy (high latitudes and Galactic plane) are likely at the origin of the 
observed discrepancy.\\ 

From Table \ref{table_cases}, the total amount of dust in the ionized phase, similar to the
dust-over-gas ratio (D/G), appears to be 3.6
and 6.4 times lower than in the atomic and molecular phases. In
particular, the PAH relative abundance with respect to BGs reveals a
decrement by a factor 2.9 and 6.6 when compared to the atomic and molecular phases, respectively.  
The high value of PAH relative abundance in the molecular phase is connected to the low X$\rm _{RF}$ which, 
in turn, is a consequence of the shielding effect from the RF induced by the higher density characterizing this phase of the gas. 
For the VSG relative abundance in the ionized phase, we do
not find the same depletion behaviour as for PAHs. The VSG relative abundance is 
in fact roughly the same in the three phases. We interpret this effect as VSGs being resiliant to the RF.

Remarkably, the spectrum of dust associated with the
ionized phase highlights more emission in the NIR domain, by a
factor of 7.5 and 18.5, compared to the molecular and atomic phases. The origin of this effect is not clear. \citet{Lu03} argue that the NIR continuum intensity is quite
linear with the strength of the aromatic features in emission. They
suggest that both this continuum and the aromatic features originate from
similar carriers. \citet{Flagey06}, using Spitzer data for 
the Galactic diffuse interstellar medium, require the presence of a continuum to explain the observations, 
but they conclude that the continuum carriers might not be the PAHs. 
What we observe in the ionized phase of the
LMC is a decrease of the PAH relative abundance and an increase of the
NIR continuum, whereas the low value of the NIR continuum intensity in the molecular phase is
associated with a high value of PAH relative abundance. Therefore our results
favour the \citet{Flagey06}'s interpretation.

\subsubsection{Case 2: typical HII regions}
In this case, the modeling seems to reveal (Figure
\ref{fig_allcases}, panel B) that emission associated with the ionized gas phase is not only  a major contributor to the total emission, but is even dominant 
with respect to the other components, i.e. atomic and molecular. Nevertheless, as in the previous section, we need to remind the reader 
that this result quite strongly depends on the electron density. At the same time, we do not think that the adopted value can 
be off by a large amount, given that it has been derived from the HII regions in the HK86 catalog. In fact, we should be wrong by a factor 5-6 in order 
for the ionized gas spectrum to match the atomic and molecular spectra at 160 $\mic$. 

For the ionized phase associated with typical HII regions, we adopt - as discussed in Section \ref{sec_mod_single} - a bluer RF input
model. Indeed, the \emph{Mathis} RF is not representative in this 
case, as the ionized gas is mostly heated by radiation from nearby stars. Since we assume 
different RFs for different gas phases, we cannot directly compare the temperature inferred from
the model. However, Figure \ref{fig_model} evidences a shift of the BG 
emission peak towards shorter wavelengths in the ionized gas phase
compared to the other phases. This result would suggest an increase of dust temperature when going from the atomic or molecular phase to
the ionized one. 
The hardness of the RF modeled with Galev is proportional to $\lambda^{-1}$. 
We have checked our results against different slopes of the input RF, for example $\lambda^{-0.5}$ and $\lambda^{-1.5}$ (see Figure \ref{fig_isrf}), 
and verified that the changes in the outputs are negligible. 

As for case 1, the ionized phase appears to be characterized by less PAHs
relative to BGs. In particular, the PAH relative abundance decreases
by a factor 4.5 to 10, compared to the relative abundance in the
atomic and molecular phases. This is likely a consequence of a higher RF
intensity with respect to the other phases. 

Furthermore, in a similar fashion to what we find for the diffuse ionized medium, the molecular phase
presents an enhancement of the PAH relative abundance. For typical HII regions, the CO data trace the
photo-dissociation region (PDR) surrounding the ionized gas, where  the local 
high density contributes to shield the PAHs from the stellar RF. 
Interestingly, the atomic phase also shows a decrease in the PAH relative
abundance compared to the molecular phase, even if the effect is less
important than for the ionized phase. The atomic phase is subject to
an attenuated stellar RF, as well as to moderate radiation from
the ISM. The combined effect of both RFs might cause the partial PAH destruction, or a decrease of 
the excitation state of the PAHs. In this latter scenario, the PAHs would be less emissive.
Overall, the results obtained for typical HII regions confirm those derived for the diffuse H$^+$. 
The VSG relative abundance is qualitatively the same in the three phases. 

As for case 1, we notice that the NIR continuum intensity is increased
in the ionized phase with respect to the atomic phase, in this case
by a factor 90, whereas the PAH relative abundance decreases. This
result corroborates the hypothesis that the NIR continuum is not correlated with the PAH abundance.

\subsubsection{Case 3: bright HII regions} \label{sec_case3_results}
In case 3, we are looking at dust emission in the vicinity of conspicuous massive
star associations. As in the previous case, dust emission in the ionized phase appears to dominate the
total emission (see Figure \ref{fig_allcases},
panels C and D), although with the caveat discussed in Section 4.1.1 and 4.1.2. 

Results derived using H$\alpha$ and radio data are in good
agreement. The lower amplitude of the spectrum associated with the ionized phase when in Equation \ref{eq_gen} we make
use of the Parkes rather than the SHASSA data is likely partly a consequence of  dust absorption
  in the SHASSA map, as already mentioned in Section \ref{sec_halpha_tracers}. 

From Table \ref{table_cases}, we notice that the RF intensity and the
dust relative abundances are in good agreement for both the
ionized gas tracers. The results show, for the ionized gas phase in particular, 
a low PAH relative abundance comparable to case 2, and an increase of the VSG relative abundance by a factor 2.2-2.5 ($\rm Y_{VSG}/Y_{BG}\simeq6.9$ in case 2 
and $\rm Y_{VSG}/Y_{BG}\simeq15.5-17.4$ in cases 3). Remarkably, these conclusions do not depend on the specific spectral shape of the RF modeled with Galev. 
The molecular phase appears again to be characterized by a high PAH relative abundance. 
We note that the PAH destruction in the ionized phase becomes more
pronounced going from case 1 to case 3, probably due to the increase of the RF intensity from the diffuse medium to bright HII regions. This 
is consistent with previous findings reported in the literature.  For instance, using ISOCAM data, \citet{Contursi07} found a lack of aromatic features carriers in the
core of the LMC HII region N4, and suggested that the high and hard RF might be
responsible for the PAH destruction. Likewise, \citet{Povich07}, identified PAH
destruction at the edge of the HII
region M17. \citet{Boulanger88} explained the decrease of the 12 $\mic$
emission in the California Nebula by grain destruction in regions
with high UV radiation. They estimated that around 80$\%$ of the
12 $\mic$ emitters could be destroyed whenever the energy
density is larger than 50 times the energy density in the solar
neighborhood. Despite this theoretical and observational evidence, we
cannot state that the RF is systematically responsible for the 
depletion of PAHs in HII regions. Indeed, several other processes could be at work, like Coulomb
destruction of PAH dications or dehydrogenation by direct
photodissociation, as suggested by \citet{Bernard94} based on the
analysis of the 3.3 $\mic$  emission and nearby continuum in the
planetary nebula BD+30 3639. Moreover, \citet{Giard94} studied the
same dust feature in the Orion bar and M17, and propose that H$^+$ chemical sputtering could cause PAH destruction.\\

In terms of temperature, as in case 2, we can deduce from Figure \ref{fig_model} the presence
of warmer dust in the ionized gas as compared to the other phases, from  
the fact that the peak of emission for the BGs is located at shorter wavelengths. 
Finally, as for case 1 and 2, the NIR continuum is higher in the ionized phase than
in the other phases.  

\subsection{Composite radiation fields}

In this section, we discuss the use of a combination of RFs, instead of a single RF, for fitting the 
SEDs corresponding to each phase of the gas (see Table \ref{table_cases_sum}). 

For case 1, the values given in Table \ref{table_cases_sum} confirm the decrease of the PAH relative 
abundance in the ionized phase, as evidenced in Section \ref{subsection_case1}. 
On the contrary, the enhancement of the PAHs in the molecular phase might suggest that these environmental 
conditions are indeed particularly favourable for their survival. For
the VSG relative abundance in the ionized phase, we find a significantly lower value with respect to what we obtain 
by adopting a single RF component (see Tables \ref{table_cases} and
\ref{table_cases_sum}). However, the VSG relative abundance is highly
dependent on the quality of the fit at FIR wavelengths and, in the ionized phase, the
model exhibits an important lack of emission at 70 $\mic$, as well as
a higher $\chi^2$, as compared to a single RF modeling. This might
indicate that the low VSG relative abundance has, for a composite RF, been underestimated. 

For case 2, the combination of the \emph{Mathis} with the \emph{4-Myr
  cluster} RFs, both for the atomic and molecular phases, generates a bluer RF than a single \emph{Mathis} RF. 
Therefore, as expected, we find a decrease in the PAH and VSG relative abundances. 
Indeed, PAHs and VSGs absorb most of their energy at short wavelengths, 
while BGs absorb energy over the whole RF spectrum. As a consequence, PAHs and VSGs absorb and re-emit more energy with a bluer RF, and to produce the same IR
flux with a bluer RF than the \emph{Mathis} RF, the model requires less
PAHs and VSGs. For the ionized phase,
the combination of \emph{4-Myr cluster} RFs gives the same PAH
relative abundance of a single \emph{4-Myr cluster} RF, whereas a significant decrease of the VSG relative abundance is observed. However, the general trend is similar to what found for 
a single RF: as in case 1, the highest PAH relative abundance is in the molecular phase, which also exhibits more VSGs with respect to BGs,
while the atomic phase shows intermediate values of PAH and VSG 
relative abundances, compared to the other gas phases.

In case 3, as for case 2, the PAH relative abundance in the atomic and
molecular phases decrease by a factor $\sim$ 2 compared to a single RF. In the ionized phase,
we still have less PAHs than in the atomic phase, by a
factor 2.2 and 2.7, for case 3a and 3b, respectively, and
by a factor between 4.4 and 5.1 compared to the molecular
phase. Results are therefore in agreement with the modeling with a
single RF. 

For the VSG relative abundance, we obtain, for the atomic phase,
different and contradictory results depending on the assumption on the RF, 
therefore preventing a clear interpretation of the results. 
In the ionized phase, the modeling confirms the enhancement of the relative VSG abundance already highlighted by a single RF. 

Overall, the results obtained assuming a composite RF substantially 
corroborate those derived with a single RF.

\subsection{Evolution of dust emission associated with different gas components}
In panels E, F and G of Figure \ref{fig_allcases}, we present the results of the analysis in a slightly different fashion, i.e. 
by combining in individual plots, each corresponding to a given gas phase, the SEDs obtained 
for all cases (1, 2 and 3). We can see that no matter which gas component 
we consider, the amplitude of the spectrum always increases going from the diffuse
medium to bright HII regions. However, for the atomic phase, the
difference in amplitude between the different cases  
is less important compared to the other 
gas components, and a rather remarkable agreement among cases is found at 8 $\mic$. As previously
noted, the absolute level of dust emission in the ionized gas is
directly proportional to $\rm n_e$, for which the adopted value changes according to
the specific environment. Therefore, the discrepancy in absolute 
amplitude of the spectra for that phase is not well constrained. On the contrary, for the atomic 
and molecular phases, we are confident that the recovered SED amplitudes are reliable, as these depend on 
the HI and CO conversion factor (Equation 1 and 2) which, respectively, is known to be constant  
for a given environment (HI), and has  been accurately determined for the
majority of the clouds (CO).

We observe prominent 8 $\mic$ emission in the atomic phase, probably associated with the 7.7 $\mic$ PAH feature. This emission feature does not 
show in the ionized phase for the case of HII
regions, likely due to PAH depletion. In comparison, the molecular phase is clearly characterized by strong 8
$\mic$ emission, as well as by significant emission at 5.8 and 12 $\mic$. 

In the FIR domain, the atomic phase presents an excess of emission at 70 $\mic$, as does the ionized phase in the diffuse medium. 
This finding is in agreement with \citet{Bernard08}.

\section{Conclusions}
\label{sec_cl}
Combining IR data with tracers of HI, H$_2$ and H$^+$, we have
evidenced, for the first time, dust emission associated with the ionized gas in the LMC. We have performed our analysis for different regimes of
the ionized gas : diffuse gas (case 1), typical HII regions
(case 2) and compact/bright HII regions (case 3). 
We have modeled the spectra associated with the atomic, molecular and
ionized phases of the gas, using different assumptions for the
radiation field, to test the robustness of our results. The results
obtained by a combination of radiations fields along the line of sight
confirm the results derived by assuming a single radiation field. 

We report a systematic warmer dust temperature in the
ionized phase compared to the atomic and molecular phases. The
inferred emissivity in the diffuse ionized gas is
2.3$\times\,10^{-26}$ cm$^2$/H at 160 $\mic$, which is lower than the
Galactic values by a factor higher than the metallicity
ratio between these 2 galaxies. This result might suggest different
properties of dust in the ionized gas of the LMC, compared to our Galaxy.
We also find, for all cases (1, 2, 3), a significant decrease of the PAH relative abundance in the ionized phase
compared to the other phases, with a larger
difference between the phases for cases 2 and 3. We interpret this result as due to PAH 
destruction, probably caused by the increased radiation field in the ionized phase, although the origin of this phenomenon is still under investigation. 

At the same time, the molecular phase appears to favor the survival of
the PAHs. In addition, when one compares bright HII regions with
typical HII regions, the ionized phase shows an
enhancement of the VSG relative abundance by more than a factor 2. We also find an important
increase of the NIR continuum in the ionized phase with respect to the other gas phases, which
does not seem to correlate directly with PAH emission. By comparing
results derived using the H$\alpha$ SHASSA map and the Parkes radio
data, the estimated extinction in H$\alpha$ for bright HII regions is
found equal to 0.51 mag (or A(V)=0.63 mag).

Finally we observe, for all gas components, a systematic increase
  of the dust emission going 
    from the diffuse medium to bright HII regions.

The origin of the destruction of the PAHs in the ionized phase (especially, in HII regions) is still not
  fully understood. Moreover, the depletion of this species is not a
  systematic effect, suggesting that changes in the PAH
  properties could also occur from one HII region to another. A
  forthcoming publication will investigate PAH depletion by means of the spectroscopic Spitzer/IRS data available in the context of the 
  SAGE-Spec Legacy survey \citep{Kemper10}. This work will allow us to
study variations of the PAH emission at small angular scale within individual HII regions. 

\acknowledgments
We thank the referee for her/his useful comments, which helped to improve
the content of the paper. We also thank M. Filipovic and A. Hughes for
helpful discussions on the Parkes radio data and on the derivation of
the thermal fraction. We acknowledge the use of the DustEM
package. The NANTEN project is based on a mutual agreement between
Nagoya University and the Carnegie Institution of Washington
(CIW). This work is financially supported in part by a Grand-in-Aid
for Scientific Research from JSPS (No. 22244014 and No. 30377931).


\newpage
\begin{thebibliography}{}
\bibitem[Bennett et al.(1992)]{Bennett92} Bennett, C. L., et al.
1992, ApJ, 396, L7
\bibitem[Bennett et al.(2003)] {Bennett03} Bennett, C. L., et al.
2003, ApJ, 583, 1
\bibitem[Beno\^it et al.(2002)]{Benoit02} Beno\^it, A., et al.
2002, Astropart. Phys., 17, 101
\bibitem[Bernard et al.(1994)]{Bernard94} Bernard, J.-P., Giard, M.,
  Normand, P., $\&$ Tiph\`ene, D.
1994, A$\&$A, 289, 524
\bibitem[Bernard et al.(2008)]{Bernard08} Bernard, J.-P., et al.
2008, ApJ, 136, 919
\bibitem[Boulanger et al.(1988)]{Boulanger88} Boulanger, F., et al.
1988, ApJ, 332, 328
\bibitem[Boulanger et al.(1996)]{Boulanger96} Boulanger, F., Abergel, A., $\&$ Bernard J.-P.
1996, A$\&$A, 312, 256
\bibitem[Compi\`egne et al.(2008)]{Compiegne08} Compi\`egne, M., et
  al.
2008, A$\&$A, 491, 797
\bibitem[Compi\`egne et al.(2010)]{Compiegne10} Compi\`egne, M., et
  al.
2011, A$\&$A, 525, 103
\bibitem[Contursi et al.(2007)]{Contursi07} Contursi, A. et al.
2007, A$\&$A, 469, 539
\bibitem[Dale et al.(2001)]{Dale01} Dale, D. A., Helou, G., Contursi, A., et al.
2001, ApJ, 549, 215
\bibitem[Davies et al.(1976)]{Davies76} Davies, R. D., Elliott, K. H.,
  $\&$ Meaburn, J.
1976, MNRAS, 81, 89
\bibitem[D\'esert et al.(1990)]{Desert90} D\'esert, F.-X., Boulanger, F., $\&$ Puget, J.-L.
1990, A$\&$A, 237, 215
\bibitem[Dickinson et al.(2003)]{Dickinson03} Dickinson, C., Davies,
  R. D., and Davis, R. J.
2003, MNRAS, 341, 369
\bibitem[Dobashi et al.(2008)]{Dobashi08} Dobashi, K., et al.
2008, A$\&$A, 484, 205
\bibitem[Everett $\&$ Churchwell(2010)]{Everett10} Everett, J. E., $\&$ Churchwell, E.
2010, ApJ, 713, 592
\bibitem[Feast(1999)]{Feast99} Feast, M.
1999, in IAU Symp. 190, New View of the Magellanic Clouds, ed. Y.-H. Chu et al. (San Francisco: ASP), 542
\bibitem[Filipovic et al.(1995)]{Filipovic95} Filipovic, M. D. et al.
1995, A$\&$AS, 111, 311
\bibitem[Flagey et al.(2006)]{Flagey06} Flagey, N. et al.
2006, A$\&$AS, 453,969
\bibitem[Fukui et al.(2008)]{Fukui08} Fukui, Y., et al. 
2008, ApJS, 178, 56
\bibitem[Gaustad et al.(2001)]{Gaustad01} Gaustad, J. E., McCullough,
  P. R., Rosing, W., $\&$ van Buren, D.
2001, PASP, 113, 1326
\bibitem[Giard et al.(1994)]{Giard94} Giard, M. et al.
1994, A$\&$A, 291, 239
\bibitem[Giardino et al.(2002)]{Giardino02} Giardino, G., et al.
2002, A$\&$A, 387, 82
\bibitem[Gordon et al.(2003)]{Gordon03} Gordon, K. D., et al.
2003, ApJ, 594, 279
\bibitem[Groves et al.(2008)]{Groves08} Groves, B., Nefs, B., $\&$
  Brandl, B.
2008, MNRAS, 391, 113
\bibitem[Haffner et al.(2009)]{Haffner09} Haffner, L. M., et al.
2009, RvMP, 81,969
\bibitem[Hauser(1993)]{Hauser93} Hauser, M. G.
1993, Back to the Galaxy, ed. S. S. Holt, $\&$ F. Verter (New-York: AIP), AIP Procc., 278, 201
\bibitem[Hoare et al.(1991)]{Hoare91} Hoare, M. G., Roche, P. F., $\&$
  Glencross, W. M.
1991, MNRAS, 251, 584
\bibitem[Hoernes et al.(1998)]{Hoernes98} Hoernes, P., Berkhuijsen,
E. M., $\&$ Xu, C.
1998, A$\&$A, 334, 57
\bibitem[Hughes et al.(2010)]{Hughes10} Hughes, A., et al.
2010, MNRAS, submitted
\bibitem[Hughes et al.(2006)]{Hughes06} Hughes, A., et al. 
2006, MNRAS, 370, 363
\bibitem[Imara $\&$ Blitz(2007)]{Imara07} Imara, N., $\&$ Blitz, L., 2007, 
ApJ, 662, 969
\bibitem[Kawamura et al.(2009)]{Kawamura09} Kawamura, A., et al.
2009, ApJS, 184, 1
\bibitem[Kemper et al.(2010)]{Kemper10} Kemper, F., et al.,
2010, PASP, 122, 683
\bibitem[Kennicutt $\&$ Hodge(1986)]{Kennicutt86} Kennicutt, R. C., $\&$ Hodge, P. W. 
1986, ApJ, 306, 130
\bibitem[Kim et al.(2003)]{Kim03} Kim, S., et al.
2003, ApJS, 148, 473
\bibitem[Kotulla et al.(2009)]{Kotulla09} Kotulla, R., Fritze, U., Weilbacher, P., $\&$ Anders, P.
2009, MNRAS, 396, 462
\bibitem[Lagache et al.(1999)]{Lagache99} Lagache, G., et al.
1999, A$\&$A, 344, 322
\bibitem[Lagache et al.(2000)]{Lagache00} Lagache, G., Haffner, L. M., Reynolds, R. J., $\&$ Tufte, S. L. 
2000, A$\&$A, 354, 247
\bibitem[van Leeuwen et al.(2007)]{Leeuwen07} van Leeuwen, F. et al.
2007, MNRAS, 379, 723
\bibitem[Lu et al.(2003)]{Lu03} Lu, N., et al.
2003, ApJ, 588, 199
\bibitem[Mathis et al.(1983)]{Mathis83} Mathis, J. S., Mezger, P. G., $\&$ Panagia, N.
1983, A$\&$A, 128,212
\bibitem[Meinert et al.(1993)]{Meinert93} Meinert, D., et al.
1993, LNP, 416, 130
\bibitem[Meixner et al.(2006)]{Meixner06} Meixner, M., et al.
2006, ApJ, 132, 2268
\bibitem[Miville-Desch\^enes $\&$ Lagache(2005)]{MamD05} Miville-Desch\^enes, M. A., $\&$ Lagache, G.
2005, ApJS, 157, 302
\bibitem[Neugebauer et al.(1984)]{Neugebauer84} Neugebauer, G., et al.
1984, ApJ, 278, L1
\bibitem[Niklas et al.(1997)]{Niklas97} Niklas, S., Klein, U., $\&$
  Wielebinski, R.
1997, A$\&$A, 322, 19
\bibitem[O$'$Donnell(1994)]{ODonnell94} O'Donnell, J. E.
1994, ApJ, 422, 158
\bibitem[Odegard et al.(2007)]{Odegard07} Odegard, N., et al.
2007, ApJ, 667, 11
\bibitem[Padoan et al.(2001)]{Padoan01} Padoan, P., Kim, S., $\&$
  Goodman, A.
2001, ApJ, 555, L33
\bibitem[Paladini et al.(2005)]{Paladini05} Paladini, R., De Zotti, G., Davies, R., $\&$  Giard, M.
2005, MNRAS 360, 1545
\bibitem[Paladini et al.(2007)]{Paladini07} Paladini, R., et al.
2007, A$\&A$ 465, 839
\bibitem[Paradis et al.(2009a)]{Paradis09a} Paradis, D. et al.
2009a, AJ, 138, 196
\bibitem[Paradis et al.(2009b)]{Paradis09b} Paradis, D. et al.
2009b, A$\&$A, 506, 745
\bibitem[Paradis et al.(2011)]{Paradis10} Paradis, D. et al.
2011, AJ, 141, 43
\bibitem[Peeters et al.(2005)]{Peeters05} Peeters, E. et al.
2005, ApJ, 620, 774
\bibitem[Planck collaboration(2011e)]{Planck11} Planck collaboration
2011e, submitted to A$\&$A, arXiv1101.2029 
\bibitem[Povich et al.(2007)]{Povich07} Povich, M. S., et al.
2007, ApJ, 660, 346
\bibitem[Reynolds et al.(1998)]{Reynolds98} Reynolds, R. J., Tufte,
  S. L., Haffner, L. M., et al.
1998, PASA, 15, 14
\bibitem[Rieke et al.(2004)]{Rieke04} Rieke, G. H., et al.
2004, ApJS, 154, 25
\bibitem[Spitzer(1978)]{Spitzer78} Spitzer, L.
1978, Physical Processes in the interstellar medium, Wiley-Interscience, New-York
\bibitem[Staveley-Smith et al.(2003)]{Staveleysmith03} Staveley-Smith, L., et al.
2003, MNRAS, 339, 87
\bibitem[Stepnik et al.(2003)]{Stepnik03} Stepnik, B., et al.
2003, A$\&$A, 398, 551
\bibitem[Watson et al.(2009)]{Watson09} Watson, C., et al.
2009, ApJ, 694, 546
\bibitem[Westerlund(1997)]{Westerlund97} Westerlund, B. E.
 1997, {\em{The Magellanic Clouds}}, New York, Cambridge Univ. Press
\bibitem[Wilson et al.(1970)]{Wilson70} Wilson, T. L., Mezger, P. G.,
  Gardner, F. F., $\&$ Milne, D. K.
1970, A$\&$A, 6, 364
\end{thebibliography}
\end{document}